**Magic Angle Spinning Effects on Longitudinal NMR Relaxation: $^{15}$N in L-Histidine**

The following article has been submitted to AIP Advances.

Armin Afrough[1)], Nichlas Vous Christensen[1,2)], Rune Wittendorff Mønster Jensen[1,3)], Dennis Wilkens Juhl[1,3)], and Thomas Vosegaard[1,3)*].

*1) Interdisciplinary Nanoscience Center, Aarhus University, Aarhus, Denmark. 2) Present address: The MR Research Centre, Department of Clinical Medicine, Aarhus University, Aarhus, Denmark. 3) Department of Chemistry, Aarhus University, Aarhus, Denmark.*

Corresponding Author: Thomas Vosegaard, tv@inano.au.dk.

**Abstract**

Solid-state magnetic resonance is a unique technique that can reveal the dynamics of complex biological systems with atomic resolution. Longitudinal relaxation is a mechanism that returns longitudinal nuclear magnetization to its thermal equilibrium by incoherent processes. The measured longitudinal relaxation rate constant however represents the combination of both incoherent and coherent contributions to the change of nuclear magnetization. This work demonstrates the effect of magic angle spinning rate on the longitudinal relaxation rate constant in two model compounds: L-histidine hydrochloride monohydrate and glycine serving as proxies for isotopically-enriched biological materials. Most notably, it is demonstrated that the longitudinal $^{15}$N relaxation of the two nitrogen nuclei in the imidazole ring in histidine is reduced by almost three orders of magnitude at the condition of rotational resonance with the amine, while the amine relaxation rate constant is increased at these conditions. The observed phenomenon may have radical implications for the solid-state magnetic resonance in biophysics and materials, especially in the proper measurement of dynamics and as a selective serial transfer step in dynamic nuclear polarization.

**Introduction**

Determining the structure and dynamics of proteins and peptides is key to understanding their function in biological systems. Solid-state Nuclear Magnetic Resonance (ssNMR) is a unique technique that can reveal such complex information by probing $^{13}$C, $^{15}$N, $^{1}$H, and $^{2}$H nuclei in biological samples through harnessing the power of magic angle spinning (MAS), cross polarization, proton decoupling, and advanced recoupling techniques. Isotopic labelling, high magnetic fields, and faster spinning rates further increase the sensitivity of ssNMR such that the structure of many proteins and peptides have been determined. The molecular structure is usually obtained from signal assignments and structural constraints while information on fast molecular motion may be obtained especially from relaxation[1, 2].

Longitudinal relaxation is an incoherent process that returns the longitudinal component of nuclear magnetization to its thermal equilibrium through internal dynamics of the molecule[3]. It has previously been pointed out that spin diffusion may influence the measured relaxation rate constants[4, 5], as the apparent longitudinal relaxation rate constant $R_1^*$ determines the combination of both *coherent* and *incoherent* contributions to the change of nuclear magnetization.[3, 6] The *incoherent*, or stochastic, contribution to signal decay contains information on the time-dependent fluctuation of bond vectors and samples the spectral density function[7]. The spectral density function is the Fourier transform of the time-correlation function that informs of the dynamics of a molecule and hence displays the intensity of dynamics at different frequencies and is largely independent of external factors like the sample spinning rate[8]. The *coherent* contribution to signal decay, however, depends on several external factors and could thus be affected by the measurement method. Understanding the dynamics of complex biological systems requires a clear understanding, and untangling of, these two contributing factors to the relaxation so that only the desired relaxation processes occur in the relaxation period and the measurement method itself does not interfere with the desired measurand.

While higher magnetic fields provide better signal separation and thus provide more details than experiments at lower fields, we here report that it may also create new challenges in the measurement and interpretation of relaxation due to rotational resonance. Rotational resonance[9] is a phenomenon that occurs when the chemical shift frequency difference $\Delta v_{ij} = v_i - v_j$ between two dipolar coupled nuclei $i$ and $j$ matches an integer multiple of the MAS rate $v_r$, i.e.,

$$|v_i - v_j| = |\Delta v_{ij}| = n v_r. \qquad (1)$$

Without chemical shift anisotropy $n$ is limited to 1 and 2; while $n$ can assume higher integer values in the presence of chemical shift anisotropy. At the rotational resonance condition, nuclear magnetization is transferred between the two nuclear sites due to the recoupling of homonuclear dipolar coupling[9]. This coherent process is independent of the dynamics of the molecule – the stochastic part of relaxation – and promotes spin diffusion.[10]

We have performed measurements of longitudinal relaxation rate constants $R_1^*$ in isotopically-enriched biological materials with $^{15}$N in two model compounds of L-histidine hydrochloride monohydrate and glycine at different spin rates ($v_r$). We have observed that the $^{15}$N longitudinal relaxation time of the two nitrogen nuclei in the imidazole ring is reduced by almost three orders of magnitude at the condition of rotational resonance with the amine.

MAS rate effects on the apparent longitudinal relaxation rate constant has far-reaching implications for ssNMR in biophysics and materials science. Some important examples include (a) the proper measurement of dynamics in peptides, (b) time-efficient ssNMR experiments[11], (c) selective transfer in dynamic nuclear polarization[12], and (d) spin-diffusion experiments[13]. Similar processes are also relevant for $^{13}$C, similar to that of $^{15}$N [14].

**Materials and Methods**

*NMR Measurements*

NMR spectra were acquired on a Bruker 950 MHz spectrometer with an Avance III HD console equipped with $^1$H-$^{13}$C-$^{15}$N-$^2$H 1.9 mm MAS probe with radiofrequency (rf) field strengths for the hard pulses and decoupling of 83.3 kHz, 28.4 kHz, and 10.9 kHz for $^1$H, $^{79}$Br, and $^{15}$N, respectively. A cross-polarization recovery pulse sequence (from Torchia[15], similar to CPXT1 in TopSpin) measured $^{15}$N longitudinal relaxation with the phase of the contact pulse and receiver inverted in alternate scans to ensure an experimental setup where the observed magnetization returns to zero. Twenty three spinning rates in the range of 5.0 to 41.3 kHz were scanned for L-histidine. SPINAL-64 proton decoupling was employed.[16] For each relaxation experiment, 64 different delay times $t_D$ in the range of 10 $\mu$s up to 10000 s were used. The FIDs were recorded using 2880 time-domain points with a spectral width of 28.8 kHz and deadtime of 6.5 $\mu$s before acquisition. Similar measurements were performed for glycine at five spinning rates of $\{6, 14.5, 23, 31.5, 40\}$ kHz.

All spectra were acquired at an internal sample temperature of 25 °C with the maximum deviation of 2.2 °C. Single-exponential $T_1^*$ relaxation time constants and spectral $T_1^*$ relaxation time distributions were obtained by in-house software.

*Materials*

Rotors were packed with polycrystalline amino acids and KBr with a ~0.8/0.2 mass ratio. Amino acids include L-histidine hydrochloride monohydrate (with 98% uniform [15]N labelling) from Cambridge Isotope Laboratories (Andover, MA, USA) and glycine (98% [15]N labeling) from Merck (Saint Louis, MO, USA). The molecular and crystalline structure of L-histidine hydrochloride monohydrate is known from X-ray and neutron studies (CCDC entry: HISTCM).[17-19] Its molecular structure consists of two groups of nearly coplanar atoms and its crystal structure belongs to the space group $P2_12_12_1$. L-histidine hydrochloride monohydrate is only referred to as L-histidine in the rest of the manuscript. [15]N in the imidazole ring of L-histidine has chemical shifts of $N^\delta$: $\delta_\delta = 189.6$ ppm (closest to branch) and $N^\varepsilon$: $\delta_\varepsilon = 176.2$ ppm. The amine [15]N chemical shift is $\delta_A = 47.3$ ppm in L-histidine and $\delta_{Gly} = 32.9$ ppm in glycine. All spectra were referenced indirectly using the chemical shift of 39.3 ppm for solid [15]$NH_4Cl$ at 25°C.[20]

*Temperature Control*

Accurate internal sample temperature control was essential to the data of this work. For all MAS rates, the sample temperature was set to 25°C by measuring the [79]Br $T_1$ relaxation of KBr, which serves as internal thermometer in the rotor.[21] Calibration curves for the difference between the set and actual temperatures were established by K[79]Br $T_1$ for different spin rates as prior information. The sample temperature was adjusted and kept with the average and maximum deviations of 1 °C and 2.2 °C from the target temperature of 25°C as assessed by the [79]Br $T_1$ relaxation time of KBr. At the highest MAS rate of 41.3 kHz, the difference between the probe set temperature and that of the internal sample was a staggering 51.5 °C. The magic angle was set by maximizing the intensity of the second sideband to the main peak of [79]Br in KBr[22].

*Numerical Simulations*

Numerical simulations were performed with SIMPSON[23-26] (version 4.2.3 for EasyNMR, https://easy.csdm.dk and 4.2.1 for Windows) on a three-spin system of [15]N in L-histidine (see Data and Codes). The dipolar coupling constants $d_{ij}$ and their respective Euler angles were calculated from the molecular structure: $d_{A\varepsilon} = -10.98$ Hz, $d_{A\delta} = -22.33$ Hz, $d_{\varepsilon\delta} = -125.98$ Hz; $\alpha_{A\varepsilon} = \alpha_{A\delta} = \alpha_{\varepsilon\delta} = 0°$; $\beta_{A\varepsilon} = 81.39°$, $\beta_{A\delta} = 59.91°$, $\beta_{\varepsilon\delta} = 123.71°$; $\gamma_{A\varepsilon} = -90.07°$, $\gamma_{A\delta} = -75.67°$, $\gamma_{\varepsilon\delta} = -117.50°$, with the Euler angles representing the orientation of the principal axes of the nuclear spin system in the crystal frame as defined by Bak et al.[23]. The chemical shift anisotropy of [15]N in L-histidine is available from single-crystal [27] and polycrystalline [28-30] samples: $\delta_{aniso,\varepsilon} = -117.1$ ppm, $\eta_\varepsilon = 0.694$, $\alpha_\varepsilon = -89.78°$, $\beta_\varepsilon = 93.44°$, $\gamma_\varepsilon = 148.90°$; $\delta_{aniso,\delta} = -127.0$ ppm, $\eta_\delta = 0.491$, $\alpha_\delta = -21.24°$, $\beta_\delta = 84.81°$, $\gamma_\delta = -30.97°$. Magnetization transfer was monitored at different MAS rates by setting the start-operator to $I_{1z}$ and detection-operator to $I_{2z}$ in SIMPSON, with nuclei 1 and 2 representing each of the different combinations of the three [15]N nuclei. The maximum signal detected at $I_{2z}$ within 1024 rotor periods (23.8-200 μs) was noted as the *effectiveness factor* $\eta$ in the range of 0 to 1; the effectiveness factor determines the possibility of magnetization exchange between two spins where $\eta = 0$ and $\eta = 1$ denote no and complete magnetization transfer, respectively, between two nuclei. Spinning rates were scanned with a resolution of 1 Hz in the MAS range of 200 Hz to 42 kHz.

**Results and Discussion**

*Longitudinal [15]N Relaxation at 22.3 T*

The imidazole $^{15}$N sites of N$^\delta$ ($\delta_\delta = 189.6$ ppm) and N$^\varepsilon$ ($\delta_\varepsilon = 176.2$ ppm) exhibit variations in $T_1^*$ (the apparent $T_1$) values over almost three orders of magnitude in the range of 6 s to 3250 s over $\nu_r \in [5\ 41.3]$ kHz (see Supplementary Material Table S1). Overall, $T_1^*(\nu_r)$ for N$^\delta$ and N$^\varepsilon$ exhibits a sigmoid-like behavior as shown in Figure 1a, while $T_1^*(\nu_r)$ for N$^A$ is approximately constant. $T_1^*$ of N$^\delta$ and N$^\varepsilon$ may be divided into three regions of MAS: Region (A) with MAS-rate-independent long $T_1^*$ for $\nu_r > 25$ kHz; Region (B) with ramp dependency of $T_1^*$ on the MAS rate for $14 \leq \nu_r < 25$ kHz; and Region (C) with shorter $T_1^*$ and complex functionality for $5 \leq \nu_r < 14$ kHz.

In Region (A) for $\nu_r \in \{30, 36, 41.3\}$ kHz $T_1^*$ is largely independent of the spin rate, leading to three distinct average $T_1^* = T_1$ values of $T_{1,\delta} = 3200 \pm 200$ s, $T_{1,\varepsilon} = 2700 \pm 350$ s, and $T_{1,A} = 1.65 \pm 0.05$ s. These values are assumed to be the true longitudinal relaxation time constants of these sites and are only influenced by incoherent effects. In Region (B) for $\nu_r \in \{14, 15, 16, 18, 20, 24\}$, $T_1^*(\nu_r)$ has a strong nearly linear dependency with slopes of 300 s/kHz and 245 s/kHz for N$^\delta$ and N$^\varepsilon$, respectively. Region (C) however demonstrates a more complex dependency on $\nu_r$ with four local minima that are better visualized in the log scale plot of Figure 1b. The amine $^{15}$N relaxation $T_{1,A}^*$ is approximately constant over the full range of spin rates, however with four local maxima that are consistent with the local minima of $T_{1,\delta}^*$ and $T_{1,\varepsilon}^*$. This semi-constant behavior, except for its local maxima, is similar to the behavior of glycine that is MAS rate independent (Figure 1b). Except at Region (A), $T_1^*$ values vary with the spin rate, and the apparent longitudinal relaxation time constant $T_1^*$ is not the same as the real longitudinal relaxation time constant $T_1$.

The local minima of the longitudinal relaxation for N$^\delta$ and N$^\varepsilon$ exactly match the rotational resonance conditions; they occur at $\nu_r = 13702$ Hz, $6851$ Hz (corresponding to the $n = 1, 2$ rotational resonance condition between N$^A$ and N$^\delta$), $\nu_r = 12412$ Hz, $6206$ Hz (corresponding to the $n = 1, 2$ rotational resonance condition between N$^A$ and N$^\varepsilon$).

Close examination of Figure 1b reveals multi-site magnetization transfers of N$^\varepsilon \leftrightarrow$ N$^\delta \leftrightarrow$ NH$_3^+$ (visible in Figure 1b at 13.702 kHz and 6.851 kHz) and N$^\delta \leftrightarrow$ N$^\varepsilon \leftrightarrow$ NH$_3^+$ (visible in Figure 1b at 12.412 kHz and 6.206 kHz) especially at the exact rotational resonance conditions. Giraud et al.[4] observed similar multi-step magnetization transfers in $^{15}$N spin diffusion experiments on a uniformly-$^{15}$N-labelled catabolite repression HPr-like protein. In their work employing 10 kHz spinning speed at 11.7 T, magnetization transfer was between linked residues in the order according to the protein sequence.

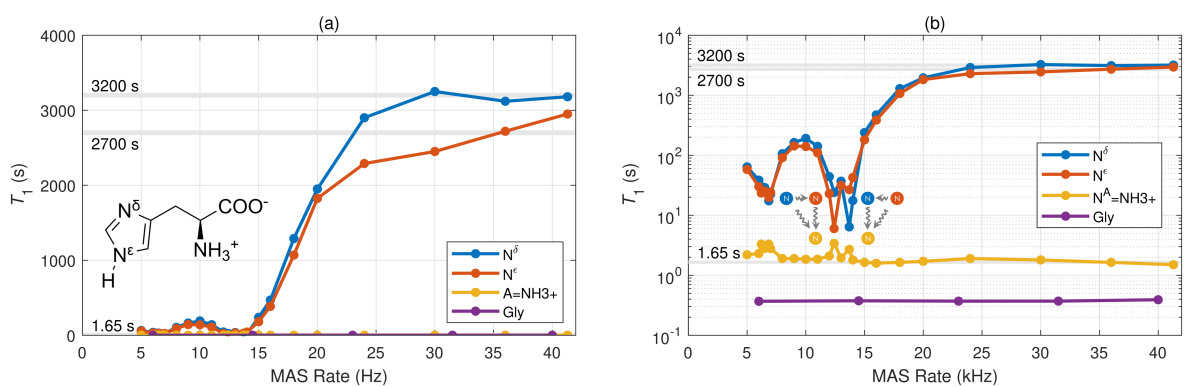

Figure 1: $^{15}$N apparent longitudinal relaxation time constants $T_1^*$ of N$^\delta$, N$^\varepsilon$, and N$^A$ ($A = $ NH$_3^+$) in L-histidine and glycine as a function of MAS rate on linear (a) and log (b) scales at 25°C. For all cases, the $T_1^*$ values are determined from monoexponential fits. For $\nu_r > 25$ kHz, $T_1^*$s are approximately independent of MAS rate and $T_1^* = T_1$. $T_{1,\delta}^*$ and $T_{1,\varepsilon}^*$ reach local minima at rotational resonance

conditions $\nu_{r,A\delta} = 13702$ Hz and $\nu_{r,A\varepsilon} = 12412$ Hz for the first rotational resonance $n = 1$ and again at $\nu_{r,A\delta} = 6851$ Hz and $\nu_{r,A\varepsilon} = 6206$ Hz for the second rotational resonance condition $n = 2$. The local maxima of $T_{1,A}^*$ ($A = \text{NH}_3^+$) match the rotational resonance conditions and corresponding minima for $T_{1,\delta}^*$ and $T_{1,\varepsilon}^*$. In contrast to L-histidine, glycine (Gly) exhibits a constant $T_{1,\text{Gly}}^* = T_{1,\text{Gly}} = 0.375$ s for all MAS rates. Lines between points are eye guides and do not have a theoretical meaning. Figure 1a and 1b show the same data with linear and logarithmic $y$-axis, respectively.

*Cross-Relaxation Rate Constants*

It is our hypothesis that the $T_1^*$ values for L-histidine may be explained by a classical magnetization-exchange model between the three sites, and with exchange/cross-relaxation rate constants being defined by incomplete averaging of the homonuclear dipole-dipole interactions. For three sites with magnetizations $I_{z,i}$, the "true" incoherent longitudinal relaxation rate constants are given by $R_{1,i} = 1/T_{1,i}$ and the cross relaxation rate constants are named $\sigma_{ij}$, $(i,j = \delta, \varepsilon, A)$. The evolution of longitudinal magnetization is governed by the matrix equation $dI/dt_D = L\,I$ with the expanded form of

$$\frac{d}{dt_D}\begin{bmatrix}I_{z,\delta}\\I_{z,\varepsilon}\\I_{z,A}\end{bmatrix} = \begin{bmatrix}-R_{1,\delta}-\sigma_{\varepsilon\delta}-\sigma_{A\delta} & \sigma_{\varepsilon\delta} & \sigma_{A\delta}\\ \sigma_{\varepsilon\delta} & -R_{1,\varepsilon}-\sigma_{\varepsilon\delta}-\sigma_{A\varepsilon} & \sigma_{A\varepsilon}\\ \sigma_{A\delta} & \sigma_{A\varepsilon} & -R_{1,A}-\sigma_{A\delta}-\sigma_{A\varepsilon}\end{bmatrix}\begin{bmatrix}I_{z,\delta}\\I_{z,\varepsilon}\\I_{z,A}\end{bmatrix} \qquad (2)$$

and solution $I(t_D) = \exp(L\,t_D)\,I(0)$; where exp is the matrix exponential and $t_D$ is the longitudinal relaxation delay. $\sigma = 0$ indicates no cross relaxation, thus no coupling between two nuclei.

A direct-search method[31] varied $\log_{10}\sigma_{A\delta}$, $\log_{10}\sigma_{A\varepsilon}$, and $\log_{10}\sigma_{\varepsilon\delta}$ in the range of $[-8, +1]$, and $\log_{10}[I_{z,\delta}(t_D = 0)]$, $\log_{10}[I_{z,\varepsilon}(t_D = 0)]$, and $\log_{10}[I_{z,A}(t_D = 0)]$ in the range of $[6.8, 7.8]$, to minimize the difference between the model values and integrals of each chemical shift peak at all longitudinal relaxation delay times. Constant $R_1$-values corresponding to $1/R_1$ values from Region (A) of Figure 1a were used for all three sites ($R_{1,\delta} = 3.145 \cdot 10^{-4}$ s$^{-1}$, $R_{1,\varepsilon} = 3.704 \cdot 10^{-4}$ s$^{-1}$, and $R_{1,A} = 0.6061$ s$^{-1}$. Optimization operations were undertaken for each MAS rate to inform on the transfer of magnetization between $^{15}$N sites (see Figure 2). Seven points (out of 69) from predicted $\sigma_{ij}$ values were regarded as outliers and were removed from results because of the bounds of optimization (see Table S2). No conditions were directly imposed on the computation of rate constants to achieve partial magnetization exchange.

Figure 2 displays the results of such parameter-estimation simulations at $\nu_r = 10$ and 12 kHz as illustrative examples. In all cases, the evolution of magnetization $I_z(t_D)$ at the three sites is matched by simulations (results not shown for all MAS rates). It was possible to match all three decay curves at all MAS rates with the single set of $R_{1,\delta}$, $R_{1,\varepsilon}$, and $R_{1,\varepsilon}$ values corresponding to $T_{1,i} = 1/R_{1,i}$ regarded as true incoherent longitudinal relaxation times and with MAS-rate-dependent cross relaxation rate constants $\sigma_{A\delta}(\nu_r)$, $\sigma_{A\varepsilon}(\nu_r)$, and $\sigma_{\varepsilon\delta}(\nu_r)$.

Experiments and simulations display several interesting features including multiexponential decay (see Figure 2c, $I_{z,A}$ at 12 kHz), and especially an increase and then decrease in magnetization (see Figure 2d, inset of $I_{z,\delta}$ at 10 kHz, bottom left). The multiexponential decay close to the first rotational resonance condition has also been observed by Kubo and McDowell[10] who recognized that the exchange of magnetization cannot be described by a single spin-diffusion time constant between the first and second rotational resonance conditions in their experiments. Although the decays are multiexponential in some cases, the dominant decay rate constant of magnetization in multiexponential analysis follows the same behavior as shown in Figure 1.

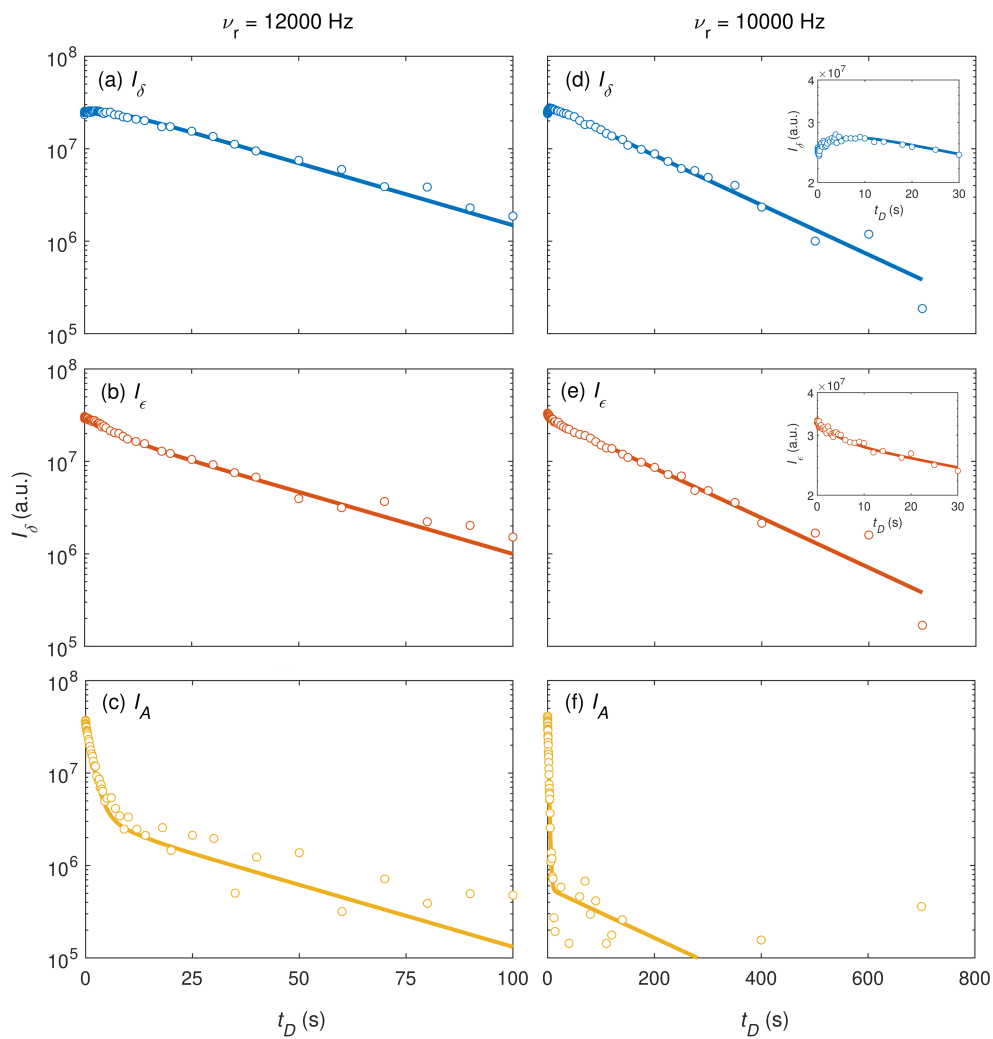

Figure 2: $^{15}$N longitudinal magnetization $I$ decay of $N^\delta$, $N^\varepsilon$, and $N^A$ (A = $NH_3^+$) in L-histidine as a function of delay time $t_D$ at two MAS rates of $\nu_r = 12$ kHz (top row) and 10 kHz (bottom row). Parameter-estimation simulations (solid line) with a single set of true longitudinal relaxation rate constants $R_{1,\delta}$, $R_{1,\varepsilon}$, and $R_{1,\varepsilon}$ matched experimental data (○). MAS-rate-dependent cross relaxation rate constants $\sigma_{A\delta}(\nu_r)$, $\sigma_{A\varepsilon}(\nu_r)$, and $\sigma_{\varepsilon\delta}(\nu_r)$ were used. Similar matches at other MAS rates are not shown.

Figure 3 shows three cross-relaxation parameters of $\sigma_{A\delta}$, $\sigma_{A\varepsilon}$, and $\sigma_{\varepsilon\delta}$ as functions of $\nu_r$ (data in Table S2). A remarkable feature of $\sigma_{A\delta}(\nu_r)$ and $\sigma_{A\varepsilon}(\nu_r)$ is that even minute cross-relaxation values are strong enough to quench $T_{1,\delta}^*$ and $T_{1,\varepsilon}^*$ and draw them closer to $T_{1,A}$. Strongest cross-relaxation parameters of $\sigma_{A\delta}(\nu_r = 13698\text{ Hz}) = 0.91$ and $\sigma_{A\varepsilon}(\nu_r = 12413\text{ Hz}) = 0.90$ coincided with the $n = 1$ rotational resonance condition; while at the second rotational resonance condition, cross-relaxation parameters are approximately an order of magnitude smaller with $\sigma_{A\delta}(\nu_r = 6849\text{ Hz}) = 0.12$ and $\sigma_{A\varepsilon}(\nu_r = 6206\text{ Hz}) = 0.076$. $\sigma_{\varepsilon\delta}$ is generally stronger than other cross-relaxation parameters, both due to the stronger dipolar coupling between these nuclei and the smaller chemical shift difference. The cross-relaxation rate constant $\sigma_{\varepsilon\delta}$ increases steadily in the direction of smaller $\nu_r$ values; spin rates around the rotational resonance condition (1290 Hz) is out of reach of the MAS probe employed in this work.

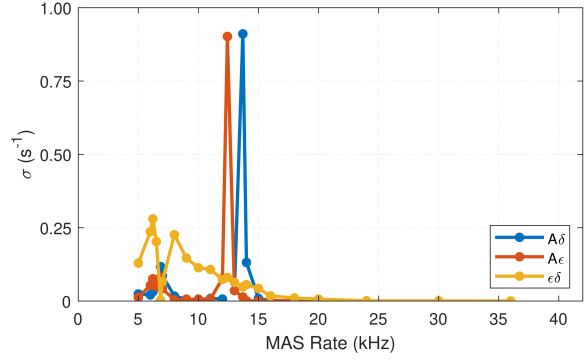

Figure 3: Calculated cross-relaxation rate constants $\sigma_{A\delta}$, $\sigma_{A\varepsilon}$, and $\sigma_{\varepsilon\delta}$ for $^{15}$N in L-histidine from parameter-estimation simulations matched to all time-domain points. Peaks of $\sigma_{A\delta}$ ($\nu_{r,n=1} = 13698$ Hz, and $\nu_{r,n=2} = 6849$ Hz) and $\sigma_{A\varepsilon}$ ($\nu_{r,n=1} = 12413$ Hz, and $\nu_{r,n=2} = 6206$ Hz) are consistent with the rotational resonance conditions. A remarkable feature of $\sigma_{A\delta}(\nu_r)$ and $\sigma_{A\varepsilon}(\nu_r)$ is that, even at spinning rates above the rotational resonance condition, small cross-relaxation values that are barely noticeable on a linear plot are strong enough to quench $T_{1,\delta}^*$ and $T_{1,\varepsilon}^*$ and draw them closer to $T_{1,A}$ (see Figure 1).

*Numerical Simulations*

The classical treatment described above does not establish a connection between the mechanism of magnetization transfer and the cross-relaxation rate constants. In relaxation experiments of this work, however, MAS-dependent recoupling of the homonuclear dipolar interaction of $^{15}$N is an obvious source of the magnetization transfer. The cross-relaxation rate constant

$$\sigma_{ij}(\nu_r) = d_{ij}^2 \, F_{ij}(\nu_r) \tag{3}$$

is proportional to the square of the dipolar coupling constant

$$d_{ij} = -\frac{\gamma_q \gamma_p \mu_0 \hbar}{4\pi r_{ij}^3} \tag{4}$$

where $r_{ij}$ is the internuclear distance between nuclei $i$ and $j$, $\gamma$ is the gyromagnetic ratio, and $\mu_0$ is the vacuum magnetic permeability. $F_{ij}(\nu_r)$ is a function of zero-quantum line shapes and hence a function of the spinning rate [10, 13, 14].

In contrast to the more phenomenological cross-relaxation rate constants introduced above, the evolution of the density matrix provides mechanistic information on the mechanisms of magnetization transfer through various field and interaction effects. We have used numerical simulations to calculate the magnetization transfer between three $^{15}$N sites in L-histidine to obtain the effectiveness factor $\eta$ as a function of MAS rate. The effectiveness factor reports on the maximum transfer of longitudinal magnetization of one nucleus to another and depends on the MAS rate. In the simulations, only chemical shift anisotropy and dipolar coupling parameters were considered for the three $^{15}$N atoms; scalar coupling parameters and protons were neglected. The effectiveness factor is analogous – although not directly comparable – to the cross-relaxation rate constant and determines the possibility of magnetization transfer between two spins where 0 and 1 denote no and complete magnetization transfer, respectively. No restrictions were imposed in numerical simulations to achieve effectiveness factors between 0 and 1.

Figure 4 demonstrates the results of numerical experiments with SIMPSON for $\eta(\nu_r)$ with $\nu_r \in [200, 42000]$ Hz (see Data and Codes). Rotational resonance was observed for at least six rotational resonance conditions (vertical lines in Figure 4) due to the presence of chemical shift anisotropy. The close similarity of Figure 4 to Figure 3 indicates the soundness of simulations and of rotational resonance recoupling of the homonuclear [15]N-[15]N dipole-dipole coupling as the source of magnetization transfer. No protons were considered in simulations, indicating that it is the direct magnetization transfer between the three [15]N sites that determines the functional dependence of longitudinal relaxation to MAS rate. Previously, it was observed that [1]H does not have effects on [15]N-[15]N and [13]C-[13]C spin diffusion as shown by deuterated uniformly-labeled solid protein[4] and peptide[14] samples (also see Agarwal et al.[32]).

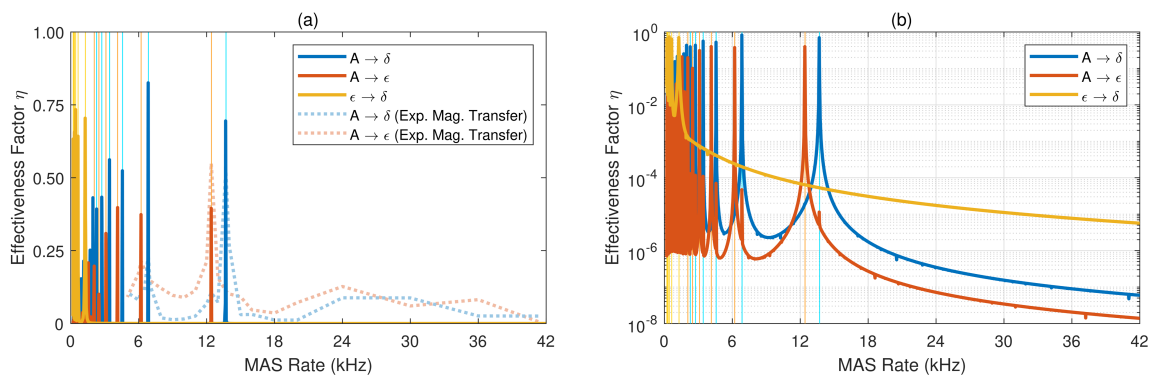

Figure 4: The effectiveness factor $\eta$ of magnetization transfer between the three [15]N nuclei in L-histidine as a function of MAS rate $\nu_r$ on linear (a) and logarithmic (b) scales. Magnetization transfer is simulated using SIMPSON at 22.3 T by considering chemical shift anisotropy and dipolar coupling for the three [15]N atoms. The peaks indicate rotational resonance conditions. The vertical lines mark expected rotational resonance conditions of the peaks with the same color. Experimental magnetization transfer (---) shows the relative ratio of the transferred longitudinal magnetization at $t_D = 4.5$ s which shows a strong correlation with the effectiveness factor.

*MAS Rate Effects in Longitudinal Relaxation ssNMR Measurements*

Andrew et al.[33] provided the first account of the effect of rotational resonance on the longitudinal relaxation. They observed an equal $T_1^*$s for two [31]P sites with distinct $T_1$ at static conditions, if the MAS rate matched that of the frequency between the two sites. Other studies have shown the surprising effects of MAS on the longitudinal relaxation rate constants. For example, Roos et al.[34] observed that moderate MAS enhances [1]H spin diffusion; while Krushelnitsky et al.[35] recognized that [15]N relaxation times in uniformly-labelled proteins can be significantly distorted by spin diffusion effects that are MAS-rate dependent. Similar dependency in [2]H longitudinal relaxation has been observed (see Cutajar et al.[36], their Tables 1 and 2).

We observed a similar behavior of longitudinal relaxation in two other series of measurements: for (1) [15]N in [15]N-L-histidine at 400 MHz and 25°C, and (2) [13]C and [15]N in [15]N-[13]C-L-histidine at 950 MHz without temperature control (both series of experiments have a lower resolution; results not shown, data available in research data). Although at much lower resolution of spinning rate and without internal sample temperature control, Fry et al.[14] independently observed similar $T_1^*$ trends for both of [13]C and [15]N in labelled Glycyl-Alanyl-Leucine 3H$_2$O (GAL) at 800 MHz. Fry et al.[14] associated this behavior not directly to the rotational resonance, but to the overlap of CSA-governed spinning sidebands with another main peak. In our model compound L-histidine, rotational resonance occurs whenever a sideband of $N^\delta$ or $N^\varepsilon$ coincides with the $NH_3^+$ resonance. Magnetization transfer is efficient if two peaks coincide, with the first rotational resonance condition being the most efficient. Although Fry et al.[14] only considered spin pairs, this picture may be extended to more complex spin systems as shown in this work.

*Effects in Common and Emerging Measurements*

Manipulation of the longitudinal relaxation time by MAS can be used to reduce experiment time. In a similar setup, band-selective pulses[11] enabled faster 2D measurements such as SOFAST-HMQC[37] by exploiting fast relaxation of specific species. Spin diffusion and ssNMR DNP techniques are two other measurements that may be affected by processes described in this work. Spin diffusion experiments are commonly employed in structural measurements and employ virtually the same pulse sequence scheme as that of the longitudinal relaxation measurement[38]. Rotational resonance recoupling provides a very specific transfer mechanism allowing long-range magnetization transfer (for example see [39]) that is a central component of structural analysis in biomolecules by ssNMR[13].

In $^{13}C$-$^{13}C$ spin diffusion measurements on $^{13}C$-labelled L-histidine, Dumez and Emsley[13] observed a complex non-monotonic influence of the spinning rate on the rate constant of magnetization transfer – analogous to the cross-relaxation rate constants measured here. They recognized that the spinning frequency dependence is associated with the chemical-shift differences between sites and is dominated by the proximity to rotational resonance. Despite their measurements at only three spinning rates of $\{10, 15, 20\}$ kHz at 16.44 T, the $^{13}C$ NMR data of Dumez and Emsley[13] show similar $\nu_r$ dependence as what we observe in our present experiments (see their Figure 4). We envision new amino acid type-selective schemes[40, 41] in ssNMR based on rotational resonance at few spinning rates could be established to complement common assignment pulse sequences. Such experiments would remove ambiguity or eliminate the need for selective labeling procedures for large protein systems[42].

The significant influence of MAS rate on the longitudinal relaxation rate constants observed here raises questions about how rotational resonance may effect magnetization transfer between $^{15}N$-$^{15}N$, $^{13}C$-$^{13}C$, and $^2H$-$^2H$[14,36] in uniformly isotope-labeled samples and hence whether all measurements are directly or indirectly affected by it. Functional dependence of the longitudinal relaxation on the spinning rate for different spin pairs guides $^{15}N$, and $^{13}C$, spin diffusion experiments by (a) optimizing magnetization transfer between a specific spin pair, (b) reducing experiment time, (c) careful selection of the magnetic field and spinning rates, and (d) promoting long-range magnetization transfer. Even in natural-abundance systems, the MAS rate affects spin diffusion in proximity to rotational resonance conditions as shown in an early study[43] (see their Figure 3).

A correct understanding of the effect of rotational resonance on the longitudinal relaxation is important also in DNP experiments – where the focus is on how to transfer magnetization to a desired nuclear site. While amine $^{15}N$ and methyl $^{13}C$ nuclei have short relaxation times and act as magnetization sinks in longitudinal relaxation measurements, they serve as entry points in DNP of biological samples[44]. For example, it was recently observed that a single methyl-nucleotide contact could be responsible for most of the DNP transfer to RNA[45]. Furthermore, it has recently been shown by Biedenbänder et al.[12] that spinning speeds close to the rotational resonance conditions expedite magnetization transfer in SCREAM-DNP[44]. A better understanding of magnetization transfer by rotational resonance offers more efficient pathways in DNP. Sweeping spinning speeds could be another strategy to expedite magnetization transfer between many sites[46].

*Perspective*

The recoupling of measurable dipolar coupling requires being almost exactly on the rotational-resonance condition – as evidenced by the very narrow resonance conditions observed in Figure 4. In fact, the rotational-resonance width experiment[47, 48] demonstrates that the width of the resonance is of similar size as that of the dipolar coupling. In this case, the dipolar couplings leading to the dramatic change in longitudinal relaxation are only in the order of 10-20 Hz. However, when the rotational resonance is to enhance subtle effects like relaxation, this requirement for proximity to the rotational-resonance condition is relieved, thereby allowing a more flexible experimental setup. The fact that proper manipulation of a weak dipolar coupling may lead to a change of three orders of magnitude for another observable parameter is very encouraging for designing new experiments to probe long-range effects in solid-state NMR. If it were possible to transfer the same types of measurements to $^1$H and measure effects of $^1$H-$^1$H homonuclear couplings down to 10-20 Hz – e.g. by deuteration and selective back-substitution of $^1$H – it would be possible to probe distances up to approximately 20 Å, thus radically changing current ways to perform solid-state NMR structural studies!

**Conclusion**

We demonstrated the effect of spin rate on the longitudinal relaxation rate constant in isotopically-enriched biological materials with $^{15}$N in two model compounds of L-histidine hydrochloride monohydrate and glycine. Rotational resonance, and mere proximity to it, significantly affected $^{15}$N relaxation. This work concludes that although not exactly at rotational resonance conditions, many ssNMR experiments may be affected by it. This phenomenon may be employed in promoting magnetization transfer between specific nuclei and utilizing shorter repetition times in experiments.

**Acknowledgements**

The use of NMR facilities at the Danish Center for Ultrahigh-Field NMR Spectroscopy funded by the Danish Ministry of Higher Education and Science (AU-1198 2010-612-181) and the Novo Nordisk Foundation (Grant no. NNF220C0075797) is acknowledged. We are grateful for financial support from the EU H2020 project PANACEA (Grant no. 101008500) and the EU Horizon-Europe project r-NMR (Grant no. 101058595).

**Supplementary Material**

Results of the fitting of the relaxation data presented in this paper are tabularized in the supplemental file named SupplementaryMaterial.pdf.

**Data and Codes**

All data and simulation input files are openly available in a public repository that issues datasets with DOIs.[49]

## Supplementary Materials

Table S1 – $^{15}$N longitudinal relaxation time constants from a single-exponential model and its standard deviation in L-histidine hydrochloride monohydrate for $N^\delta$ ($\delta_{N^\delta} = 189.6$ ppm), $N^\varepsilon$ ($\delta_{N^\varepsilon} = 176.2$ ppm), and $N^\delta$ ($\delta_{NH_3^+} = 47.3$ ppm) at 22.3 T.

| Exp. No. | $v_r$ (kHz) | $T_{1,N^\delta}$ (s) | $SD[T_{1,N^\delta}]$ (s) | $T_{1,N^\varepsilon}$ (s) | $SD[T_{1,N^\varepsilon}]$ (s) | $T_{1,NH_3^+}$ (s) | $SD[T_{1,NH_3^+}]$ (s) |
|---|---|---|---|---|---|---|---|
| 408 | 41.300 | 3180 | 170 | 2950 | 115 | 1.50 | 0.05 |
| 412 | 36.000 | 3120 | 120 | 2720 | 100 | 1.65 | 0.05 |
| 418 | 30.000 | 3250 | 220 | 2450 | 130 | 1.80 | 0.05 |
| 420 | 24.000 | 2900 | 270 | 2290 | 180 | 1.9 | 0.1 |
| 426 | 20.000 | 1950 | 60 | 1825 | 75 | 1.71 | 0.07 |
| 430 | 18.000 | 1290 | 45 | 1070 | 40 | 1.64 | 0.06 |
| 432 | 16.000 | 470 | 20 | 386 | 13 | 1.59 | 0.05 |
| 435 | 15.000 | 239 | 18 | 181 | 8 | 1.64 | 0.06 |
| 440 | 14.000 | 17.6 | 1.2 | 42.5 | 3.5 | 1.8 | 0.1 |
| 444 | 13.698 | 6.4 | 0.5 | 26.7 | 1.8 | 2.7 | 0.3 |
| 442 | 13.000 | 37 | 2 | 31.5 | 1.2 | 1.95 | 0.15 |
| 444 | 12.413 | 24 | 1.5 | 6 | 0.3 | 3.4 | 0.3 |
| 447 | 12.000 | 44 | 3 | 22.9 | 1.1 | 2.1 | 0.15 |
| 448 | 11.000 | 141 | 9 | 110 | 4 | 1.85 | 0.08 |
| 450 | 10.000 | 191 | 14 | 140 | 7 | 1.85 | 0.06 |
| 452 | 9.000 | 163 | 8 | 143 | 4.5 | 1.90 | 0.06 |
| 453 | 8.000 | 106 | 8 | 91 | 3 | 1.94 | 0.08 |
| 455 | 7.000 | 22.5 | 1.5 | 24.2 | 0.8 | 2.8 | 0.25 |
| 457 | 6.849 | 17.4 | 0.8 | 19.7 | 0.6 | 3.3 | 0.3 |
| 461 | 6.500 | 29 | 2 | 25.3 | 0.8 | 2.8 | 0.24 |
| 460 | 6.206 | 30 | 2 | 23.5 | 0.9 | 3.3 | 0.3 |
| 458 | 6.000 | 38.5 | 2.5 | 30 | 1 | 2.28 | 0.15 |
| 462 | 5.000 | 64 | 4 | 58 | 2 | 2.23 | 0.12 |

Table S2 – Cross-relaxation rates for [15]N longitudinal relaxation in L-histidine hydrochloride monohydrate at 22.3 T. Omitted data were regarded as outliers with results obtained out of bounds of parameter-estimation optimization.

| MAS (kHz) | $\sigma_{N^\varepsilon-NH_3^+}(s^{-1})$ | $\sigma_{N^\delta-NH_3^+}(s^{-1})$ | $\sigma_{N^\delta-N^\varepsilon}(s^{-1})$ |
|---|---|---|---|
| 41.300 | | | |
| 36.000 | | 0.000011 | 0.00031 |
| 30.000 | | 0.000034 | 0.00057 |
| 24.000 | | 0.00011 | 0.0010 |
| 20.000 | 0.000010 | 0.00036 | 0.0066 |
| 18.000 | 0.00028 | 0.00075 | 0.011 |
| 16.000 | 0.0022 | 0.0019 | 0.018 |
| 15.000 | 0.00065 | 0.0091 | 0.043 |
| 14.000 | 0.00050 | 0.13 | 0.056 |
| 13.698 | 0.014 | 0.91 | 0.048 |
| 13.000 | 0.036 | 0.035 | 0.064 |
| 12.413 | 0.90 | | 0.081 |
| 12.000 | 0.077 | 0.0070 | 0.073 |
| 11.000 | 0.0070 | 0.0091 | 0.11 |
| 10.000 | 0.0067 | 0.0053 | 0.11 |
| 9.000 | 0.0062 | 0.0066 | 0.15 |
| 8.000 | 0.0041 | 0.017 | 0.23 |
| 7.000 | 0.042 | 0.087 | 0.066 |
| 6.849 | 0.081 | 0.12 | 0.0030 |
| 6.500 | 0.050 | 0.048 | 0.20 |
| 6.206 | 0.076 | 0.032 | 0.28 |
| 6.000 | 0.054 | 0.021 | 0.24 |
| 5.000 | 0.013 | 0.024 | 0.13 |